\documentclass[aps,prd,reprint,superscriptaddress,nofootinbib,showpacs]{revtex4-2}

\usepackage{setspace}  % alters the baseline spacing on load
\usepackage[section]{placeins}  % floats cannot cross a section boundary
\usepackage{microtype}  % justification and line breaking

\usepackage[T1]{fontenc}
\usepackage{amsmath,amssymb,mathtools}

\usepackage{booktabs}  % \toprule, \midrule, \bottomrule

\usepackage{xcolor}  % \textcolor
\usepackage[normalem]{ulem}  % \sout  (normalem: keep \emph as italic)
\usepackage[hidelinks]{hyperref}  % links carry no color box; load after the rest

\begin{document}

\title{A window for the mixed same-sign signature in the type-II seesaw models}

\author{Cheng-Wei Chiang}
\affiliation{Department of Physics and Center for Theoretical Physics, National Taiwan University, Taipei, Taiwan 10617, R.O.C.}
\affiliation{Physics Division, National Center for Theoretical Sciences, Taipei, Taiwan 10617, R.O.C.}

% --------------------------------------------------------------
\begin{abstract}

The mass splitting between the doubly and singly charged members of the Higgs triplet of the type-II seesaw model is usually treated as a free parameter in the collider literature.  Electroweak loops fix it to $767$ to $841$~MeV over the mass range considered here once the quartic coupling $\lambda_4$ is set to zero at the renormalization scale $\overline\mu = M_{H^{\pm\pm}}$.  Moving that scale by a factor of two in either direction moves the loop value by $208$~MeV at $200$~GeV, falling to $42$~MeV at $1$~TeV.  At the reference splitting of $852$~MeV, the transition $H^{\pm\pm} \to H^\pm W^{\pm(*)}$ takes $40\%$ ($7\%$) of the total width at the crossover triplet vacuum expectation value for a doubly charged Higgs mass of $200$~GeV ($500$~GeV).  That splitting is also where the partonic counting of that rate fails.  Evaluated from hadronic data, the counting overestimates the rate by a factor $1.46 \pm 0.05$ at a splitting of $852$~MeV and underestimates it by a factor $4.87 \pm 0.09$ at $200$~MeV.  Whether the mixed same-sign topology survives is then a condition on $\lambda_4$.  The same channel moves the two published exclusion boundaries in opposite directions, by $-1.4\%$ and $+24.5\%$, at the level of branching ratios and at fixed experimental acceptance.

\end{abstract}

\maketitle

%%%%%%%%%%%%%%%%%%%%%%%%%%%%%%%%%%%%%%%%%%%%%%%%%%
\section{Introduction}
\label{sec:intro}
%%%%%%%%%%%%%%%%%%%%%%%%%%%%%%%%%%%%%%%%%%%%%%%%%%

The type-II seesaw mechanism is an economical renormalizable way to give neutrinos their observed tiny mass without introducing a right-handed partner.  Here, a single complex $SU(2)_L$ triplet $\Delta$ of hypercharge $Y = 2$ is added to the Higgs sector, giving the so-called Higgs triplet model (HTM)~\cite{Schechter:1980gr,Cheng:1980qt,Magg:1980ut,Lazarides:1980nt,Mohapatra:1980yp}.  The mechanism has attractive collider phenomenology because of a direct connection between the properties of the additional Higgs bosons and neutrino mass.

One remarkable feature of the model is the existence of a doubly charged Higgs boson $H^{\pm\pm}$ carrying two units of electric charge, a charge carried by no Standard Model (SM) particle.  Its production via the Drell-Yan (DY) process at hadron colliders is determined solely by its electroweak charges~\cite{FileviezPerez:2008jbu,Chakrabarti:1998qy,Fuks:2019clu}.  On the other hand, how $H^{\pm\pm}$ decays depends on the model parameters, one of which is the triplet vacuum expectation value (VEV), denoted by $v_\Delta$.

Since the neutrino mass $M_\nu = \sqrt2\, Y_\Delta v_\Delta$, with $Y_\Delta$ the Yukawa coupling, the oscillation data constrain only the product and leave $v_\Delta$ free.  Without a mass splitting with the singly charged partner, $H^{\pm\pm}$ has two primary decay modes.  The leptonic mode produces two same-sign leptons through the same Yukawa matrix that generates the neutrino mass, at a rate falling as $v_\Delta^{-2}$.  The gauge mode produces two same-sign $W$ bosons, at a rate rising as $v_\Delta^{2}$.

Published experimental searches assume one of the two extremes.  The dilepton analysis takes the leptonic modes to saturate the total width, while the diboson analysis assumes the same for the gauge modes~\cite{ATLAS:2022pbd,ATLAS:2021jol}.  Between them lies a triplet VEV for each given mass, where the two rates cross and neither assumption holds.

Around that crossover VEV, a pair-produced sample can be divided into three same-sign topologies in the decay products: four leptons, four $W$ bosons, and the mixed one with leptons on one side and $W$ bosons on the other.  The mixed topology is the largest of the three at that point because of the combinatorics.  It is also the only one of the three that carries both a lepton-number-violating vertex and a gauge vertex in the same event.  That makes it the natural object to search for in this regime, as proposed in Refs.~\cite{delAguila:2013yaa,delAguila:2013mia,Babu:2022ycv,Bolton:2024thn}.

The coverage lost around that point is structural rather than statistical.  A search excludes a point only where the predicted rate exceeds the published upper limit.  Let $\sigma_{\rm DY}$ be the theoretical Drell-Yan cross section for $H^{++}H^{--}$ and $\sigma_{95}$ the $95\%$ confidence level upper limit the search places on it.  The condition then reads $\sigma_{\rm DY}\,{\rm BR}^2 \ge \sigma_{95}$, which at a given mass constrains the branching ratio alone, ${\rm BR} \ge \sqrt{\sigma_{95}/\sigma_{\rm DY}}$.

The dilepton analysis therefore reaches up from small $v_\Delta$ only as far as the leptonic branching ratio remains above that value.  The diboson analysis reaches down from large $v_\Delta$ only as far as the gauge one does.  Both fall short near the crossover VEV, where the crossover branching ratio of each mode is about one half and the rate into either pure topology is down by a factor of four.  On a logarithmic axis in $v_\Delta$, the range left between them sits symmetrically about the crossover VEV, since the two boundaries are related by reflection in it whenever the two searches demand the same threshold branching ratio.

That range follows from the two analyses.  This paper shows that cascade decay would widen the range asymmetrically by an amount not considered in the published works.  However, a range bounded on both sides and symmetric about the crossover VEV presumes that both searches cover the same mass.

This is not the case in the current searches based on the $139$~fb$^{-1}$ dataset of the CERN Large Hadron Collider (LHC).  The diboson analysis excludes nothing above $340$~GeV, while the published dilepton limit is tabulated only from $400$~GeV upward, so no single mass is bounded by both.  Over $200$ to $340$~GeV, only the diboson analysis applies, and it excludes $v_\Delta$ above its boundary.  Over $400$ to $1000$~GeV, only the dilepton analysis applies, and it excludes $v_\Delta$ below its boundary, leaving every larger value uncovered up to the few GeV permitted by the $\rho$ parameter.  The two-sided range requires the mass ranges of the two searches to overlap, which needs more data than $139$~fb$^{-1}$, as discussed in Sec.~\ref{sec:limits}.

This paper is organized as follows.  Section~\ref{sec:model} gives a short review of the HTM and introduces the notation.  Section~\ref{sec:splitting} obtains the mass splitting between the doubly and singly charged Higgs bosons.  It is given first as the exact tree relation in terms of the quartic coupling $\lambda_4$, and then as the electroweak loop value it takes at $\lambda_4 = 0$, with the calculation given in Appendix~\ref{sec:app-loop}.  Section~\ref{sec:cascade} discusses the cascade transition $H^{\pm\pm} \to H^\pm W^{\pm(*)}$ that the splitting opens, whose rate is independent of $v_\Delta$ and grows as $\Delta_{\rm split}^5$.  Section~\ref{sec:hadronic} argues that the free quark counting of that rate is incorrect and replaces it by one built from hadronic data, with two evaluations set out in Appendix~\ref{sec:app-routes}.  Section~\ref{sec:window} turns the result into the range of $\lambda_4$ over which the mixed topology survives at a given mass.  Section~\ref{sec:limits} takes into account the same channel in the published dilepton and diboson exclusions, showing that the two boundaries move in opposite directions.  Section~\ref{sec:summary} summarizes our findings.

%%%%%%%%%%%%%%%%%%%%%%%%%%%%%%%%%%%%%%%%%%%%%%%%%%
\section{The Higgs triplet model}
\label{sec:model}
%%%%%%%%%%%%%%%%%%%%%%%%%%%%%%%%%%%%%%%%%%%%%%%%%%

In the HTM, the triplet field of hypercharge $Y = 2$ can be written in the adjoint representation as
\begin{equation}
\Delta =
\begin{pmatrix}
\delta^+/\sqrt2 & \delta^{++} \\
\delta^0 & -\delta^+/\sqrt2
\end{pmatrix} ,
\label{eq:let-triplet}
\end{equation}
with the electric charge fixed by $Q = T^3 + Y/2$ and $T^a = \sigma^a/2$.  The triplet couples to the rest of the model through its gauge, Yukawa and scalar potential terms.  The kinetic term
\begin{align}
\begin{split}
\mathcal{L}_{\rm kin} &= {\rm Tr}\!\left[\left(D_\mu \Delta\right)^\dagger D^\mu \Delta\right] 
~, \\
D_\mu \Delta &= \partial_\mu \Delta + i g \left[W^a_\mu T^a, \Delta\right] + i g' B_\mu \Delta 
~,
\end{split}
\label{eq:let-kin}
\end{align}
carries the gauge coupling.  The Yukawa term
\begin{equation}
\mathcal{L}_Y = \left(Y_\Delta\right)_{ij} L_i^T C\, i\sigma_2\, \Delta\, L_j + \text{h.c.} 
~,
\label{eq:let-yuk}
\end{equation}
with $Y_\Delta$ symmetric, violates lepton number by two units and generates the Majorana neutrino mass $M_\nu = \sqrt2\, Y_\Delta v_\Delta$.  Writing $\Phi = (\phi^+, \phi^0)^T$ for the SM doublet, the scalar potential
\begin{align}
\begin{split}
V \supset {} & \mu\, \Phi^T {\rm i}\sigma_2 \Delta^\dagger \Phi + {\rm h.c.} \\
& {} + \lambda_1 \left(\Phi^\dagger\Phi\right) {\rm Tr}\!\left(\Delta^\dagger\Delta\right)
+ \lambda_2 \left[{\rm Tr}\!\left(\Delta^\dagger\Delta\right)\right]^2 \\
& {} + \lambda_3 {\rm Tr}\!\left[\left(\Delta^\dagger\Delta\right)^2\right]
+ \lambda_4 \, \Phi^\dagger \Delta \Delta^\dagger \Phi
~,
\end{split}
\label{eq:let-lam4}
\end{align}
alongside the doublet terms in the SM.  The coefficient $\mu$ breaks lepton number softly and generates $v_\Delta$.  Minimizing the potential with $M_\Delta$ the triplet mass gives the standard induced VEV of the type-II seesaw~\cite{Schechter:1980gr,Cheng:1980qt,Magg:1980ut,Lazarides:1980nt,Mohapatra:1980yp}, $v_\Delta \simeq \mu v_\Phi^2/(\sqrt2 M_\Delta^2)$.  Setting $\mu = 0$ restores lepton number conservation.  A small $\mu$ is therefore stable against radiative corrections, and $v_\Delta$ remains a free parameter rather than being fixed by the rest of the potential.

Of the four quartic couplings, only $\lambda_4$ splits the triplet members at ${\cal O}(v_\Phi^2)$, as seen in the tree spectrum derived in Refs.~\cite{Melfo:2011nx,Aoki:2011pz,Arhrib:2011uy}.  Substituting the doublet VEV gives $\lambda_4 \Phi^\dagger \Delta \Delta^\dagger \Phi \to \lambda_4 (v_\Phi^2/2)\left(\Delta\Delta^\dagger\right)_{22}$.  This picks out one entry of the triplet matrix and weights the components \emph{linearly} by their third isospin component, rendering the three states equally spaced at tree level.

In contrast, $\lambda_1$ multiplies the trace ${\rm Tr}(\Delta^\dagger\Delta)$ and treats the three states alike.  The coupling $\lambda_2$ multiplies the square of that trace and does not separate them at tree level or even at one loop.  The coupling $\lambda_3$ is quartic in $\Delta$ and separates them only at order ${\cal O}(v_\Delta^2)$.

With the VEV definitions $\langle \phi^0\rangle = v_\Phi/\sqrt2$ and $\langle \delta^0\rangle = v_\Delta/\sqrt2$, the weak scale is $v^2 = v_\Phi^2 + 2 v_\Delta^2 = (246.22~{\rm GeV})^2$.  The triplet shifts the $\rho$ parameter by $\rho - 1 \simeq -2 v_\Delta^2/v^2$, which keeps $v_\Delta$ below a few GeV.  With the oblique parameters $S = 0.008 \pm 0.071$ and $T = 0.021 \pm 0.055$ at $U = 0$~\cite{PDG2026} and $\rho_0 = 1 + \alpha(M_Z) T$, we quote $\rho_0 - 1 = (1.6 \pm 4.3)\times 10^{-4}$.  Since the triplet VEV drives $\rho$ below one, the constraint is the lower edge of that interval, and $v_\Delta \le [-(\rho_0-1) v^2/2]^{1/2}$ gives $v_\Delta \le 2.8$~GeV at one standard deviation and $4.6$~GeV at two.

The physical states are $H^{\pm\pm}$, $H^\pm$, $H^0$, $A^0$ and the $125$-GeV Higgs boson $h$.  The doubly charged state is $\delta^{\pm\pm}$ exactly, since nothing else in the theory carries two units of electric charge.  The mixing of $\delta^\pm$ with $\phi^\pm$ is $\sqrt2\, v_\Delta/v_\Phi$, which stays at or below $10^{-5}$ for $v_\Delta \lesssim 1.7 \times 10^{-3}$~GeV.  In what follows, we work with $M_{H^{\pm\pm}}$ and $v_\Delta$ as the two parameters and set the mixings to zero.

%%%%%%%%%%%%%%%%%%%%%%%%%%%%%%%%%%%%%%%%%%%%%%%%%%
\section{Mass splitting}
\label{sec:splitting}
%%%%%%%%%%%%%%%%%%%%%%%%%%%%%%%%%%%%%%%%%%%%%%%%%%

The phenomenology below is controlled by the mass splitting inside the triplet:
\begin{equation}
\Delta_{\rm split} = M_{H^{\pm\pm}} - M_{H^\pm}
~,
\label{eq:let-dsplit}
\end{equation}
taken positive, so that the doubly charged state is the heavier of the two.  Its value is set by two sources, the scalar potential at tree level and the electroweak gauge loops.  The lower gap is taken against $A^0$ rather than $H^0$.  Diagonalizing the singly charged and the CP-odd mass matrices generated by Eq.~\eqref{eq:let-lam4}, the two adjacent gaps in mass squared are~\cite{Melfo:2011nx,Aoki:2011pz,Arhrib:2011uy}
\begin{align}
\begin{split}
M_{H^{\pm\pm}}^2 - M_{H^+}^2 
&= 
-\frac{\lambda_4 v_\Phi^2}{4} - \sqrt2 \mu v_\Delta + \left( \frac{\lambda_4}{2} - \lambda_3 \right) v_\Delta^2
~, \\
M_{H^+}^2 - M_{A^0}^2 
&= 
-\frac{\lambda_4 v_\Phi^2}{4} - \sqrt2 \mu v_\Delta - \frac{\lambda_4}{2}\, v_\Delta^2
~.
\end{split}
\label{eq:let-gaps-exact}
\end{align}
The term $-\lambda_4 v_\Phi^2/4$ is identical in both lines.  The third terms are quadratic in $v_\Delta$ and are the only ones that differ between the two lines.  They break this uniformity by the difference $\left(\lambda_4 - \lambda_3\right) v_\Delta^2$.  The second term, $-\sqrt2\,\mu\,v_\Delta$, is again common to both gaps and lowers them by the same amount, so the equal spacing is preserved.  It nevertheless enters each gap, and therefore shifts $\Delta_{\rm split}$ itself.

Once $\Delta_{\rm split} \ll M_{H^{\pm\pm}}$, one has $\Delta_{\rm split} \simeq (M_{H^{\pm\pm}}^2 - M_{H^+}^2)/(2 M_{H^{\pm\pm}})$.  The vacuum relation $\mu = \sqrt2 M_\Delta^2 v_\Delta/v_\Phi^2$ of Sec.~\ref{sec:model} turns the $\mu$ term into $\sqrt2 \mu v_\Delta = 2 M_\Delta^2 v_\Delta^2/v_\Phi^2$.  Therefore, it contributes to the mass splitting by $\sim M_{H^{\pm\pm}}\left(v_\Delta/v_\Phi\right)^2$ , which reaches at most $17$~eV for $M_{H^{\pm\pm}} = 1$~TeV and $v_\Delta = 10^{-3}$~GeV.
The quadratic terms become $\lambda v_\Delta^2/(2M_{H^{\pm\pm}})$ and reach at most $3$~eV, at $200$~GeV and for a coupling $\lambda$ of ${\cal O}(1)$.  Both corrections are negligible compared with the splittings of several hundred MeV obtained below.  Therefore, the potential contributes to Eq.~\eqref{eq:let-dsplit} effectively via the single term $-\lambda_4 v_\Phi^2/(8 M_{H^{\pm\pm}})$.

The gauge boson loops contribute in addition and raise the state with larger $|Q|$.  Together, the mass splitting is
\begin{equation}
\Delta_{\rm split}(M_{H^{\pm\pm}}) 
= 
\Delta_{\rm split}^{\rm loop}(M_{H^{\pm\pm}}) - \frac{\lambda_4 v^2}{8 M_{H^{\pm\pm}}} 
~.
\label{eq:let-family}
\end{equation}
The loop term is $767$~MeV at $M_{H^{\pm\pm}} = 200$~GeV rising to $841$~MeV at $1$~TeV.  These values follow from the finite-mass one-loop function of Ref.~\cite{Cirelli:2005uq} in modified minimal subtraction, with $\lambda_4(\overline\mu) = 0$ at the renormalization scale $\overline\mu = M_{H^{\pm\pm}}$.  This condition defines a reference point at which the entire mass splitting is of loop origin.  Gauge loop renormalization makes $\lambda_4$ nonzero at other scales, and only the sum in Eq.~\eqref{eq:let-family} is independent of $\overline\mu$.

%%%%%%%%%%%%%%%%%%%%
\begin{table}[tbp]
\centering
\caption{The loop term $\Delta_{\rm split}^{\rm loop}$ of Eq.~\eqref{eq:let-family} at three renormalization scales, with $\lambda_4(\overline\mu)$ set to zero at $\overline\mu = M_{H^{\pm\pm}}$.  The last column gives the swing across that range, rounded to the closest integer.  Its product with the mass is constant, $\simeq 41.7$~GeV$^2$.  The last row is the heavy multiplet limit, Eq.~\eqref{eq:let-closed}, which is scale-independent.}
\label{tab:loop-scale}
\medskip
\begin{tabular}{crcccc}
\toprule
$M_{H^{\pm\pm}}$ & \multicolumn{4}{c}{$\Delta_{\rm split}^{\rm loop}(\overline\mu)$ [MeV]} & swing \\
$[$GeV$]$ & $\overline\mu =$ & $M_{H^{\pm\pm}}/2$ & ~$M_{H^{\pm\pm}}$~ & $2M_{H^{\pm\pm}}$ & [MeV] \\
\midrule
$200$  && $663$ & $767$ & $872$ & $208$ \\
$350$  && $752$ & $812$ & $871$ & $119$ \\
$500$  && $784$ & $826$ & $868$ & $83$  \\
$700$  && $805$ & $835$ & $865$ & $60$  \\
$1000$  && $820$ & $841$ & $862$ & $42$  \\
\midrule
$\infty$ && $852$ & $852$ & $852$ & $0$  \\
\bottomrule
\end{tabular}
\end{table}
%%%%%%%%%%%%%%%%%%%%

The choice of scale matters more than this mass dependence.  Table~\ref{tab:loop-scale} gives the splitting at $\overline\mu = M_{H^{\pm\pm}}/2$, $M_{H^{\pm\pm}}$ and $2M_{H^{\pm\pm}}$.  The column for $\overline\mu = M_{H^{\pm\pm}}$ changes by $74$~MeV between $M_{H^{\pm\pm}} = 200$~GeV and $1$~TeV.  The swing $\delta\Delta_{\rm split}^{\rm loop}$ across the scale range, by contrast, would be canceled by the shift in $\lambda_4$ that accompanies the change of renormalization scale.  The product of the swing with the mass is a constant, $\simeq 41.7$~GeV$^2$.  By Eq.~\eqref{eq:let-family}, a shift $\delta\lambda_4$ moves $\Delta_{\rm split}$ by $-\delta\lambda_4 v^2/(8 M_{H^{\pm\pm}})$.  Therefore,
\begin{equation}
\delta\lambda_4 = \frac{8}{v^2} \times 41.7~{\rm GeV}^2 = 0.0055 
~,
\label{eq:let-loopswing}
\end{equation}
the same at every mass.  The splitting is therefore fixed once $\lambda_4$ is specified at a stated scale.  The residual freedom lies within $\lambda_4$, the one parameter that the paper already treats as unknown.

The loop function tends to $852$~MeV in the heavy multiplet limit\footnote{Ref.~\cite{FileviezPerez:2008jbu} quotes $540$~MeV for this same pair, taking it from Ref.~\cite{Cirelli:2005uq}.  That entry is in fact the splitting Ref.~\cite{Cirelli:2005uq} gives between the singly charged and the neutral state, corresponding to our evaluation of $529$~MeV.  For the doubly charged against the singly charged pair, the same closed form returns $852$~MeV.}, where it takes the closed form
\begin{equation}
\frac{\alpha_{\rm em}\left[ \left(Q^2 - Q'^2\right) M_W + Y \left(Q - Q'\right) M_Z \right]}
{2\left(1 + \cos\theta_W\right)}
\label{eq:let-closed}
\end{equation}
for the gap between two adjacent members~\cite{Dimopoulos:1990kc,Mizuta:1992ja,Thomas:1998wy,Cirelli:2005uq}, evaluated with the tree-level $\alpha_{\rm em} = g^2 \sin^2\theta_W/4\pi = 1/132.05$.\footnote{The Thomson value would give $821$~MeV and $\alpha_{\rm em}(M_Z)$ would give $880$~MeV.  We quote $852$~MeV throughout as a fixed reference value rather than as the model's value at any particular mass.}  The triplet has two adjacent gaps standing in the fixed ratio
\begin{equation}
\frac{M_{H^{\pm\pm}} - M_{H^\pm}}{M_{H^\pm} - M_{H^0}}
= \frac{3M_W + 2M_Z}{M_W + 2M_Z} = 1.612 
~,
\label{eq:let-gapratio}
\end{equation}
independent of $\alpha_{\rm em}$.  The same evaluation that gives $852$~MeV for the upper gap therefore gives $529$~MeV for the lower one.

According to Eq.~\eqref{eq:let-family}, a particular value of $\lambda_4$ cancels the loop contribution exactly:
\begin{equation}
\lambda_4^{\rm cancel} = \frac{8 M_{H^{\pm\pm}} \Delta_{\rm split}^{\rm loop}}{v^2}
\simeq 0.020
\label{eq:let-cancel}
\end{equation}
at $M_{H^{\pm\pm}} = 200$~GeV and rising to $0.11$ at $1$~TeV.  For $\lambda_4$ above $\lambda_4^{\rm cancel}$, the ordering inverts and the doubly charged state becomes the lightest of the triplet.  In that case the cascade decay discussed below is closed by kinematics.  At the crossover VEV, the mixed fraction then returns to the $1/2$ counted above.  Away from that point, the two are unequal and the product is smaller.  The lepton-number-violating three-body mode $H^{\pm\pm} \to W^\pm \ell^\pm \bar\nu$, to be discussed in Sec.~\ref{sec:window}, survives the inversion and keeps the sum slightly below unity, which removes $2.4\%$ of that $1/2$ at $M_{H^{\pm\pm}} = 1$~TeV and nothing at $200$~GeV.  The mixed topology therefore cannot be predicted without specifying $\lambda_4$.

%%%%%%%%%%%%%%%%%%%%%%%%%%%%%%%%%%%%%%%%%%%%%%%%%%
\section{Cascade channel}
\label{sec:cascade}
%%%%%%%%%%%%%%%%%%%%%%%%%%%%%%%%%%%%%%%%%%%%%%%%%%

We restrict our attention to the regime where $\Delta_{\rm split} \ll M_W$ and $\Delta_{\rm split} (M_{H^{\pm\pm}}) > 0$, and consider a cascade transition inside the triplet:
\begin{equation}
H^{\pm\pm} \;\longrightarrow\; H^\pm\, W^{\pm(*)}
~,
\label{eq:let-cascade}
\end{equation}
coming from the derivative term of the triplet kinetic energy and carrying the gauge coupling with no dependence on $v_\Delta$.  Unlike the two primary channels, its rate is set by the splitting in Eq.~\eqref{eq:let-dsplit} rather than by the triplet VEV.  Far below the $W$ threshold, the exchanged $W$ is off shell over its entire range, and the propagator can be expanded in $q^2/M_W^2 \le \Delta_{\rm split}^2/M_W^2 \sim 10^{-4}$ and replaced by $-1/M_W^2$.  The transition then reduces to a contact interaction of Fermi strength $g^2/M_W^2$.

Suppose the final state fermion masses are set to zero, and the splitting is small compared to $M_{H^{\pm\pm}}$.  The available phase space is then that of a three-body decay whose only scale is $\Delta_{\rm split}$.  Using the off-shell integral set up in Ref.~\cite{Akeroyd:2011zza}, the partonic decay width\footnote{This formula includes nine channels: the three lepton pairs $e^+\nu_e$, $\mu^+\nu_\mu$ and $\tau^+\nu_\tau$, and the two quark pairs $u\bar d$ and $c\bar s$, each counted in three colors.  All nine are open only in the massless limit.  At the splittings of interest here, the $\tau$ and the charm channels are forbidden, and are treated in Sec.~\ref{sec:hadronic}.} is~\cite{FileviezPerez:2008jbu}:
\begin{equation}
\Gamma_{\rm parton} \equiv \Gamma\!\left(H^{\pm\pm} \to H^\pm W^{\pm *}\right)
\to
\frac{3 g^4 \Delta_{\rm split}^5}{80 \pi^3 M_W^4}
~.
\label{eq:let-fifth}
\end{equation}
The fifth power makes the cascade width strongly sensitive to the splitting.  Writing $c$ for the branching fraction of the cascade decay, each of the three same-sign topology fractions falls by the same factor of $(1-c)^2$.  Since the factor is common, the cascade does not disfavor the mixed topology relative to the other two, but reduces the yield of all three equally.  The mixed topology therefore remains observable only while $c$ is small, and by the fifth power this bounds $\Delta_{\rm split}$ from above.

When the leptonic and diboson widths become equal, both branching ratios fall to about $1/2$, the signal in either search is reduced by a factor of 4, and the excluded mass drops accordingly.  This is already visible in the published limits.  A recent scan at zero splitting reports an exclusion increasing from $420$~GeV at large $v_\Delta \sim 10^{-3}$~GeV to $955$~GeV at small $v_\Delta \sim 10^{-5}$~GeV~\cite{Ashanujjaman:2021txz}.  Both channels are controlled by $v_\Delta$ in opposite senses, and both widths are known results~\cite{FileviezPerez:2008jbu,Melfo:2011nx,Han:2015hba}, rederived below in the conventions used here.

The triplet Yukawa coupling is fixed by the neutrino mass relation $Y_\Delta = M_\nu/(\sqrt2\, v_\Delta)$, with $M_\nu$ the neutrino mass matrix in the basis where the charged lepton masses are diagonal.  The width into one unordered flavor pair is $\left|(M_\nu)_{ij}\right|^2 M_{H^{\pm\pm}}/[8\pi (1+\delta_{ij}) v_\Delta^2]$.  The factor $(1+\delta_{ij})^{-1}$ is the identical particle factor carried by the three same flavor pairs.  Summing the six unordered pairs collapses the flavor structure onto a trace, so that the mixing angles and the Majorana phases cancel and only the mass eigenvalues survive, giving
\begin{equation}
\Gamma_{\ell\ell} = \frac{M_{H^{\pm\pm}}\, \mathcal{T}}{16\pi\, v_\Delta^2}
~, \qquad
\mathcal{T} \equiv {\rm Tr}\!\left(M_\nu^\dagger M_\nu\right) = \sum_i m_i^2
~,
\label{eq:let-Gll}
\end{equation}
with the charged lepton masses neglected.

The diboson coupling comes from the triplet kinetic term, which contains $-(g^2 v_\Delta/\sqrt2)\, H^{\pm\pm} W^\mp_\mu W^{\mp\mu}$, with the vertex factor twice the coefficient by combinatorics.  Summing the nine polarization combinations and halving the phase space for the two identical $W$ bosons, we obtain
\begin{equation}
\Gamma_{WW} = \frac{g^4 v_\Delta^2 M_{H^{\pm\pm}}^3}{64\pi\, M_W^4}\, P
~,
\label{eq:let-GWW}
\end{equation}
where $P \equiv \sqrt{1-4r}\left(1 - 4r + 12 r^2\right)$, with $r \equiv M_W^2/M_{H^{\pm\pm}}^2$, is the phase space factor, vanishing at threshold and tending to unity for $M_{H^{\pm\pm}} \gg M_W$.

For each mass, the two channels are equally strong at one triplet VEV, called the crossover VEV and denoted by $v_\Delta^{\rm cross}$.  We write $\Gamma_{\ell\ell} = A/v_\Delta^2$ and $\Gamma_{WW} = C v_\Delta^2$, with $A = M_{H^{\pm\pm}}\mathcal{T}/16\pi$ and $C = g^4 M_{H^{\pm\pm}}^3 P/(64\pi M_W^4)$.  Setting them equal gives
\begin{equation}
\left(v_\Delta^{\rm cross}\right)^4
= \frac{A}{C}
= \frac{4\, \mathcal{T} M_W^4}
{g^4 M_{H^{\pm\pm}}^2 \sqrt{1-4r}\left(1 - 4r + 12 r^2\right)}
~.
\label{eq:let-vcross}
\end{equation}
The common decay width at the crossover VEV scales as $M_{H^{\pm\pm}}^2 \sqrt{\mathcal{T}}\,\sqrt{P}$.  The factor $\sqrt{P}$ cannot be neglected: $P$ runs from $0.397$ at $M_{H^{\pm\pm}} = 200$~GeV to $0.857$ at $500$~GeV.  It converts the naive scaling $(500/200)^2 = 6.25$ into the value $9.18$ quoted in Sec.~\ref{sec:hadronic}.  We take the normal ordering with a lightest neutrino mass of $0.03$~eV throughout, which gives $\sum_i m_i = 0.120$~eV.  As an example, $v_\Delta^{\rm cross} \simeq 6.9 \times 10^{-5}$~GeV at $M_{H^{\pm\pm}} = 500$~GeV.

%%%%%%%%%%%%%%%%%%%%%%%%%%%%%%%%%%%%%%%%%%%%%%%%%%
\section{Hadronic calculation}
\label{sec:hadronic}
%%%%%%%%%%%%%%%%%%%%%%%%%%%%%%%%%%%%%%%%%%%%%%%%%%

Equation~\eqref{eq:let-fifth} includes nine massless fermion channels with equal weight.  At the splittings that Eq.~\eqref{eq:let-family} produces, a few hundred MeV to about a GeV, that counting is qualitatively incorrect.  The virtual $W$ cannot produce free quark pairs, but light mesons instead.  We therefore compute the rate from hadronic data in two independent ways: from the spectral functions measured in $\tau$ decay~\cite{Davier:2013sfa} and from a sum over exclusive meson channels.

\begin{figure}[tbp]
\centering
\includegraphics[width=\columnwidth]{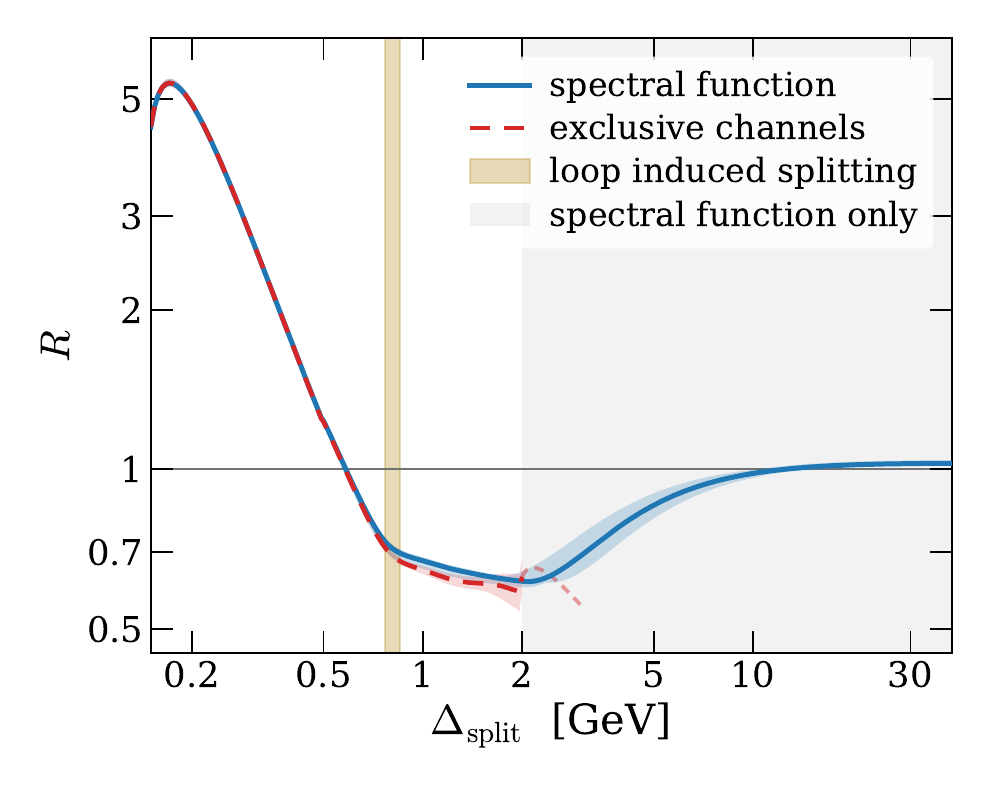}
\caption{The ratio $R$ of Eq.~\eqref{eq:let-Rdef}, as a function of the mass splitting for $M_{H^{\pm\pm}} = 500$~GeV.  The blue solid curve uses the measured spectral function, continued by the perturbative series above $\sqrt s = 1.55$~GeV.  The red dashed curve follows the exclusive channels, drawn paler from $2$ to $3$~GeV and not drawn beyond.  The light bands give each method's own uncertainty.  The gray region marks the splittings at which the exclusive channel list can no longer be completed, leaving only the spectral function above $3$~GeV.  The ochre vertical band indicates the loop splitting, from $767$~MeV at $M_{H^{\pm\pm}} = 200$~GeV up to the heavy multiplet limit $852$~MeV.  The horizontal line at unity is where the hadronic and the partonic rates agree.}
\label{fig:let-r9}
\end{figure}

Writing $\Gamma_{\rm hadron}$ as the cascade rate evaluated by the hadronic calculation, we plot in Fig.~\ref{fig:let-r9} the ratio
\begin{equation}
R = \frac{\Gamma_{\rm hadron}}{\Gamma_{\rm parton}} 
~.
\label{eq:let-Rdef}
\end{equation}
In the sub-GeV mass splitting regime, the charged weak current has matrix elements to $\pi$, $\pi\pi$, $K\pi$, and so on.  Summing over all of them gives the hadronic rate.  The partonic rate is the leading term of the operator product expansion of that same quantity, the expansion of the current correlator in inverse powers of $s$.  Its unit operator gives the free quark term, and its dimension-four condensates enter as $m_q\left\langle \overline{q}q\right\rangle/s^{2}$ and $\left\langle GG\right\rangle/s^2$~\cite{Braaten:1991qm}.  Quark-hadron duality holds only if the integration range $0 < s < \Delta_{\rm split}^2$ is sufficiently wide to average over the resonances it contains.

At $\Delta_{\rm split} = 400$~MeV, that range is $0.16$~GeV$^2$, compared to $m_\tau^2 = 3.16$~GeV$^2$ over which duality is tested in $\tau$ decay.  It is a factor of twenty narrower and contains only the pion and the two-pion threshold rather than a resonance to average over.  The partonic rate there is the leading term of an expansion that has broken down.  The two methods agree to within $2.4\%$ of their mean over the range where both apply, $200$~MeV to $2$~GeV.  The ratio
\begin{align}
\begin{split}
R(852~\text{MeV}) &= 0.684 \pm 0.022 
~, \\
R(540~\text{MeV}) &= 1.112 \pm 0.024 
~,
\end{split}
\label{eq:let-R9}
\end{align}
and reaching $4.874 \pm 0.090$ at $200$~MeV.

%%%%%%%%%%%%%%%%%%%%
%%%%%%%%%%%%%%%%%%%%

For sufficiently small splitting, the hadronic rate is larger than the partonic rate.  The partonic rate counts a three-body final state, the $H^\pm$ and a quark pair, growing as $\Delta_{\rm split}^5$.  The single pion channel is the two-body decay $H^{\pm\pm} \to H^\pm \pi^\pm$, growing as $f_\pi^2 \Delta_{\rm split}^3$.  Their ratio is
\begin{equation}
R_\pi = \frac{5\pi^2}{3}\,\frac{\left|V_{ud}\right|^2 f_\pi^2}{\Delta_{\rm split}^2} 
~.
\label{eq:let-Rpi}
\end{equation}
The two-body phase space is therefore enhanced relative to the three-body one by $\Delta_{\rm split}^{-2}$ as the splitting decreases.  Below the two-pion threshold, $m_{\pi^\pm} + m_{\pi^0} = 274.6$~MeV, the pion is the only state the current can produce.  When the mass splitting is larger than about $600$~MeV, the hadronic rate becomes smaller for the opposite reason.  The nine channels (three lepton pairs and two quark pairs in three colors) are not all open.  The $\tau\nu$ (charm) channel is forbidden below a splitting of $m_\tau = 1.777$~GeV ($2$~GeV for charm).  The light quark continuum has not yet saturated the partonic counting either.

The ratio $R$ compares the sum of the channels that are actually open against the nine-channel law, which weights all nine equally.  At a few hundred MeV, the electron and muon pairs are open and the tau pair is not, and those two pairs appear in both the numerator and the denominator.  Counted as massless, they would carry $2/9$ of the nine-channel rate, so that equality of the hadronic and partonic rates means that the pion should supply the rest to make $R_\pi = 1 - 2/9 = 7/9$.  Solving Eq.~\eqref{eq:let-Rpi} at that value gives in closed form the splitting at which $R = 1$:
\begin{equation}
\Delta_\times = \pi \sqrt{15/7}\, \left|V_{ud}\right| f_\pi = 583~\text{MeV} 
~,
\label{eq:let-cross}
\end{equation}
with $f_\pi = 130.2$~MeV and $|V_{ud}| = 0.97431$ from the unitarity fit of Ref.~\cite{PDG2026}, the direct determinations there giving $0.97367$ and moving $\Delta_\times$ by $0.4$~MeV.  Sending $m_\pi \to 0$ in the full calculation moves $\Delta_\times$ by only $1.5\%$.

The two scales that control the comparison share no common input.  The loop splitting is $\alpha_{\rm em}$ times an electroweak mass [$\simeq 112.55$~GeV according to Eq.~\eqref{eq:let-closed}], with no strong interaction in it.  In contrast, $\Delta_\times$ of Eq.~\eqref{eq:let-cross} carries $f_\pi$ and no $\alpha_{\rm em}$, its only electroweak inputs being $|V_{ud}|$ and the lepton thresholds.  The two scales are nevertheless close in the sub-GeV regime.  Whether the partonic calculation overestimates or underestimates the cascade decay rate therefore depends on which side of $\Delta_\times$ the loop splitting lies on.  As shown in Fig.~\ref{fig:let-r9}, the reference splitting is above $\Delta_\times$ and gives $R = 0.68$.

%%%%%%%%%%%%%%%%%%%%
\begin{figure*}[tbp]
\centering
\begin{minipage}{0.485\textwidth}
\centering
\includegraphics[width=\textwidth]{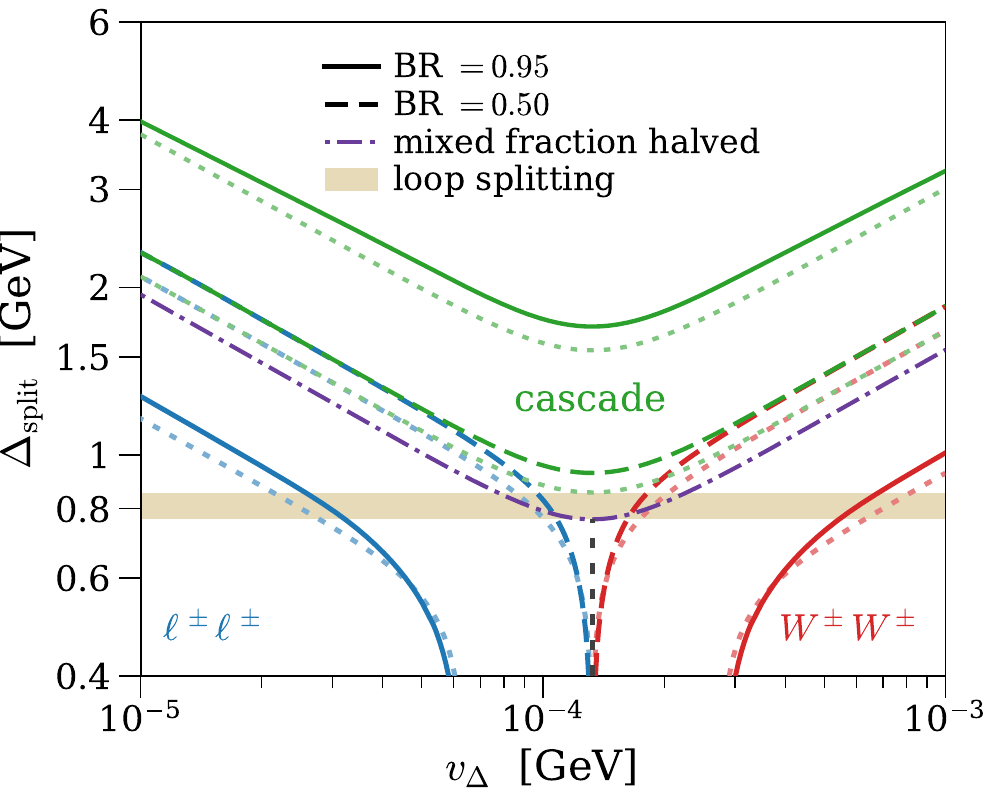}\\[2pt]
(a) $M_{H^{\pm\pm}} = 200$~GeV
\end{minipage}
\hspace{0.014\textwidth}
\begin{minipage}{0.485\textwidth}
\centering
\includegraphics[width=\textwidth]{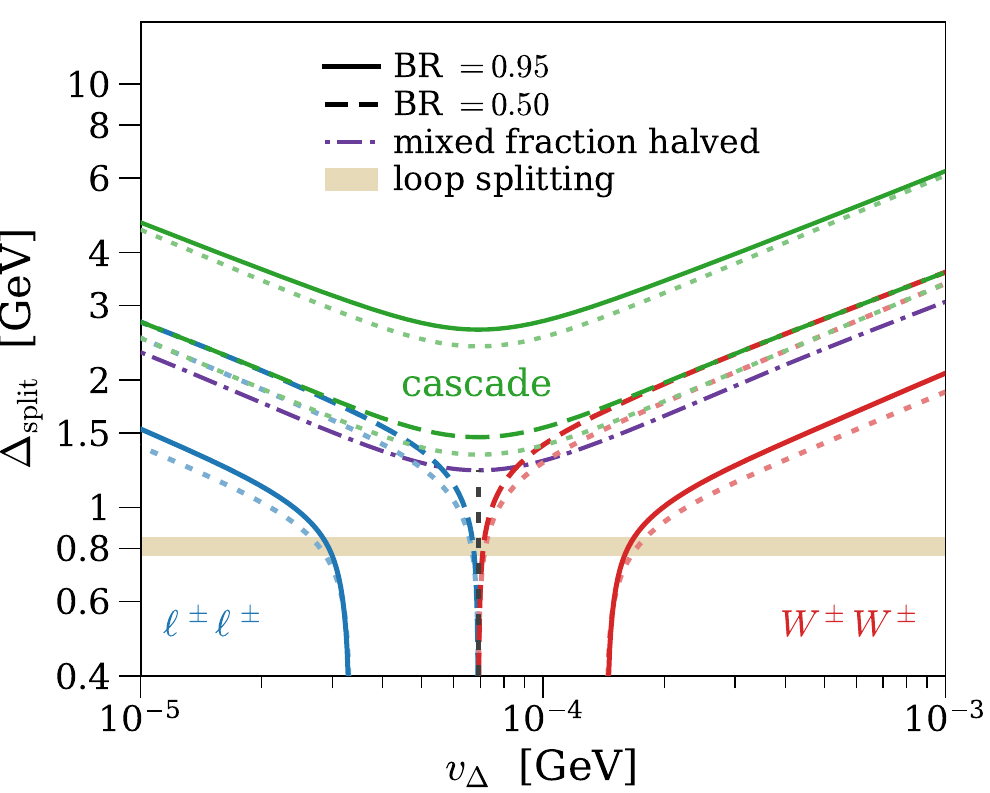}\\[2pt]
(b) $M_{H^{\pm\pm}} = 500$~GeV
\end{minipage}
\caption{Mass splitting against triplet VEV, at (a) $M_{H^{\pm\pm}} = 200$~GeV and (b) $500$~GeV.  Color gives the decay mode, labeled on the plane: blue for $\ell^\pm\ell^\pm$, red for $W^\pm W^\pm$, green for the cascade.  Line style gives the level, solid for a branching ratio of $0.95$ and dashed for $0.50$.  Each same-sign contour appears twice: the plain curve is computed with the hadronic cascade rate, taken as the mean of the two methods of Appendix~\ref{sec:app-routes}, and the dotted curve, in a paler shade of the same color, uses the partonic rate of Eq.~\eqref{eq:let-fifth}.  Purple dash-dotted: where the mixed fraction $2\,{\rm BR}(\ell\ell)\,{\rm BR}(WW)$ is half its value at zero splitting, computed with the hadronic rate.  Ochre band: the loop splitting band of Fig.~\ref{fig:let-r9}.  Short gray vertical rule: the crossover VEV, $v_\Delta^{\rm cross} \simeq 1.3 \times 10^{-4}$~GeV in (a) and $6.9 \times 10^{-5}$~GeV in (b), where $\Gamma_{\ell\ell} = \Gamma_{WW}$.  Neutrino masses are NuFIT 6.0~\cite{Esteban:2024eli} normal ordering, the variant without Super-Kamiokande atmospheric data, with a lightest mass of $0.03$~eV.}
\label{fig:let-phase}
\end{figure*}
%%%%%%%%%%%%%%%%%%%%

Fig.~\ref{fig:let-phase}(a) shows the effect in the plane of the triplet VEV and the splitting for $M_{H^{\pm\pm}} = 200$~GeV.  At the crossover VEV and the reference splitting, the cascade takes $40\%$ of the total width, $c = 0.401$, so that the mixed fraction is $(1-c)^2/2 \simeq 0.179$ in contrast with $1/2$ when the cascade is absent.  The purple contour, on which that fraction has halved, runs \emph{inside} the ochre loop splitting band, so switching the quartic coupling off is by itself sufficient to halve the mixed topology.  Above $\Delta_\times$, where $R < 1$, each hadronic contour lies above its partonic counterpart, a larger splitting being needed to reach the same fraction.  The shift scales as $R^{-1/5}$, from the fifth power of Eq.~\eqref{eq:let-fifth}.

Fig.~\ref{fig:let-phase}(b) shows that the effect is much weaker when $M_{H^{\pm\pm}} = 500$~GeV.  The cascade width does not depend on the mass, while $\Gamma_{\ell\ell} = \Gamma_{WW}$ at the crossover VEV grows with it.  Thus, their sum is $9.18$ times larger than at $M_{H^{\pm\pm}} = 200$~GeV, with the cascade taking only $7\%$ of the total width rather than $40\%$, leaving a mixed fraction of $0.434$.  The loop splitting band lies well below every cascade contour there, whereas at $200$~GeV the contour at ${\rm BR}({\rm cascade}) = 0.50$ lies only a few MeV above the top of the band.

Let $v_{\ell\ell}(b)$ and $v_{WW}(b)$ be the triplet VEVs at which the two branching ratios reach a common value $b$ at a given mass and splitting.  Using $A/C = \left(v_\Delta^{\rm cross}\right)^4$ from Eq.~\eqref{eq:let-vcross}, we have
\begin{equation}
v_{\ell\ell}(b)\; v_{WW}(b) = \left(v_\Delta^{\rm cross}\right)^2
~,
\label{eq:let-mirror}
\end{equation}
independent of $b$ and the cascade width $\Gamma_{\rm casc}$.  The two contours are thus mirror images about $v_\Delta^{\rm cross}$ on a logarithmic axis, as noted in Sec.~\ref{sec:intro}.  The two boundaries therefore move apart by reciprocal factors, and the range left uncovered widens without its center moving, as to be seen in Sec.~\ref{sec:limits}.

The cascade formula used in the most complete recent scan is the same partonic integral as Eq.~\eqref{eq:let-fifth}, continued down to $\Delta_{\rm split} = 0.1$~GeV~\cite{Ashanujjaman:2021txz}.  For example, between the two splittings quoted in the literature, $540$ and $852$~MeV, that integral gives the naive fifth power $(852/540)^5 = 9.78$.  The hadronic evaluation gives $9.78 \times R(852~{\rm MeV})/R(540~{\rm MeV}) \simeq 6.01 \pm 0.09$ instead.  Every branching ratio computed near the loop-induced splitting follows from the partonic width of Eq.~\eqref{eq:let-fifth} rescaled by $R$.

%%%%%%%%%%%%%%%%%%%%%%%%%%%%%%%%%%%%%%%%%%%%%%%%%%
\section{Window for mixed topology}
\label{sec:window}
%%%%%%%%%%%%%%%%%%%%%%%%%%%%%%%%%%%%%%%%%%%%%%%%%%

Because $\Delta_{\rm split}$ is adjustable, it is commonly scanned.  Some set collider limits at $0$, $\pm 10$ and $\pm 30$~GeV~\cite{Ashanujjaman:2021txz}; close the channel by making the doubly charged state the lightest of the triplet~\cite{Antusch:2018svb}; or argue that a splitting below about a GeV leaves the cascade width under the other two~\cite{BhupalDev:2018tox}.  This paper answers the last: the loops supply $852$~MeV, and the cascade is not small.  All three avoid the one splitting the model produces when $\lambda_4$ is turned off.

At the crossover VEV of a given mass, the total width is the smallest and the mixed topology the largest\footnote{The total width is stationary at ${\rm d}\Gamma/{\rm d}v_\Delta = -2A/v_\Delta^3 + 2 C v_\Delta = 0$, that is at $v_\Delta^4 = A/C$, the same condition as $\Gamma_{\ell\ell} = \Gamma_{WW}$.  Since $\Gamma_{\ell\ell}\Gamma_{WW} = AC$ is independent of $v_\Delta$, the mixed fraction $2\Gamma_{\ell\ell}\Gamma_{WW}/\Gamma^2$ peaks where $\Gamma$ is the smallest.}.  The cascade fraction $c = \Gamma_{\rm casc}/(\Gamma_{\rm casc} + \Gamma_{\ell\ell} + \Gamma_{WW})$ is therefore largest there as well.

\begin{figure}[tbp]
\centering
\includegraphics[width=1.05\columnwidth]{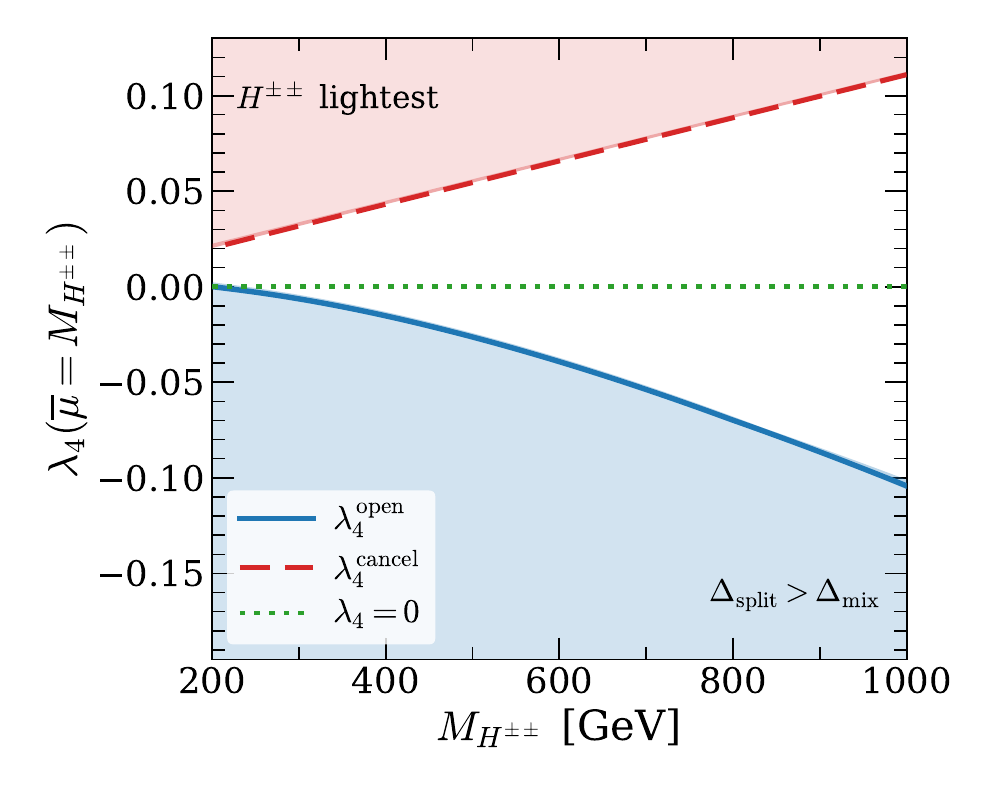}
\caption{Quartic coupling $\lambda_4$ against the doubly charged Higgs mass.  The blue solid is $\lambda_4^{\rm open}$, where $\Delta_{\rm split}$ reaches $\Delta_{\rm mix}$.  Each of the two hadronic methods gives its own cascade rate and therefore its own $\Delta_{\rm mix}$, and the curve uses the average of the two.  The red dashed curve gives $\lambda_4^{\rm cancel}$ of Eq.~\eqref{eq:let-cancel}, on the same complete one-loop splitting.  The blank region between the two curves is where the mixed topology survives, with $H^{\pm\pm}$ being the heaviest of the triplet.}
\label{fig:let-lambda4}
\end{figure}

Let $\Delta_{\rm mix}$ be the splitting where the cascade decay causes the mixed topology rate to be halved.  Since $2\,{\rm BR}(\ell\ell)\,{\rm BR}(WW) \propto (1-c)^2$, this implies $c = 1 - 1/\sqrt2 \simeq 0.293$.  Let $\lambda_4^{\rm open}(M_{H^{\pm\pm}})$ be the coupling where $\Delta_{\rm split}$ of Eq.~\eqref{eq:let-family} reaches it, the companion of $\lambda_4^{\rm cancel}$ in Eq.~\eqref{eq:let-cancel}.  The mixed topology survives for
\begin{equation}
\lambda_4^{\rm open}\!\left(M_{H^{\pm\pm}}\right) < \lambda_4 < \lambda_4^{\rm cancel}\!\left(M_{H^{\pm\pm}}\right)
~.
\label{eq:let-interval}
\end{equation}
The interval has a lower end because, according to Eq.~\eqref{eq:let-family}, the splitting increases as $\lambda_4$ drops.  Therefore, the requirement $\Delta_{\rm split} < \Delta_{\rm mix}$ imposes a lower bound on $\lambda_4$.  Above $\lambda_4^{\rm cancel}$ the splitting changes sign, $H^{\pm\pm}$ becomes the lightest of the triplet, and the cascade is closed by kinematics.  Figure~\ref{fig:let-lambda4} shows both curves against the doubly charged Higgs mass.

%%%%%%%%%%%%%%%%%%%%
%%%%%%%%%%%%%%%%%%%%

Equating $\Delta_{\rm split}$ of Eq.~\eqref{eq:let-family} to $\Delta_{\rm mix}$ gives the lower end in closed form, $\lambda_4^{\rm open} = 8 M_{H^{\pm\pm}}\left(\Delta_{\rm split}^{\rm loop} - \Delta_{\rm mix}\right)/v^2$.  It requires $\lambda_4 > 4.1 \times 10^{-5}$ at $M_{H^{\pm\pm}} = 200$~GeV and turns negative at $201$~GeV.  Here $\Delta_{\rm mix}$ is obtained by solving $c = 1 - 1/\sqrt2$ at the crossover VEV for the splitting, with $\Gamma_{\rm casc}$ taken from the hadronic calculation of Sec.~\ref{sec:hadronic}.  It grows with the mass faster than $\Delta_{\rm split}^{\rm loop}$, which only runs from $767$ to $841$~MeV over the same mass range, so the two cross at $201$~GeV and the closed form changes sign there.  Above $201$~GeV, the lower bound is therefore negative, so $\lambda_4 = 0$ already keeps the mixed topology.  At $200$~GeV, the two hadronic methods do not settle the sign, giving $+9.5 \times 10^{-5}$ and $-1.4 \times 10^{-5}$, a spread of $\pm 5.4 \times 10^{-5}$ about the mean that straddles zero.  The running of the coupling is given by~\cite{Chao:2006ye}
\begin{equation}
\frac{{\rm d}\lambda_4}{{\rm d}\log\overline\mu} = \frac{3 g^2 g'^2 Y}{8\pi^2} \simeq 4.0 \times 10^{-3}
~,
\label{eq:let-running}
\end{equation}
taken from the one-loop renormalization group equations of this potential, after neglecting terms that are proportional to $\lambda_4$ itself or to the triplet Yukawa coupling.  So the gauge term dominates the running in this regime.  It moves $\lambda_4$ by $9.1 \times 10^{-3}$ over a factor of ten in scale, which is over $200$ times the requirement at $200$~GeV.  Whether the lower bound is met is therefore settled by the scale at which $\lambda_4$ is quoted rather than by its magnitude.  Throughout this paper, it is $\lambda_4(\overline\mu = M_{H^{\pm\pm}})$.

The interval is not empty: $4.1 \times 10^{-5} < \lambda_4 < 2.0 \times 10^{-2}$, the upper end being $\lambda_4^{\rm cancel}$ at the lowest mass, keeps the mixed topology at every mass from $200$~GeV to $1$~TeV.  Both ends are set by the lowest mass, since $\lambda_4^{\rm open}$ falls from $4.1 \times 10^{-5}$ to $-0.10$ at $1$~TeV while $\lambda_4^{\rm cancel}$ rises from $0.020$ to $0.11$.  The window demands $\lambda_4 > \lambda_4^{\rm open}$, a condition on the potential that analyses with a free splitting never face.

%%%%%%%%%%%%%%%%%%%%
\begin{figure}[tbp]
\centering
\includegraphics[width=1.05\columnwidth]{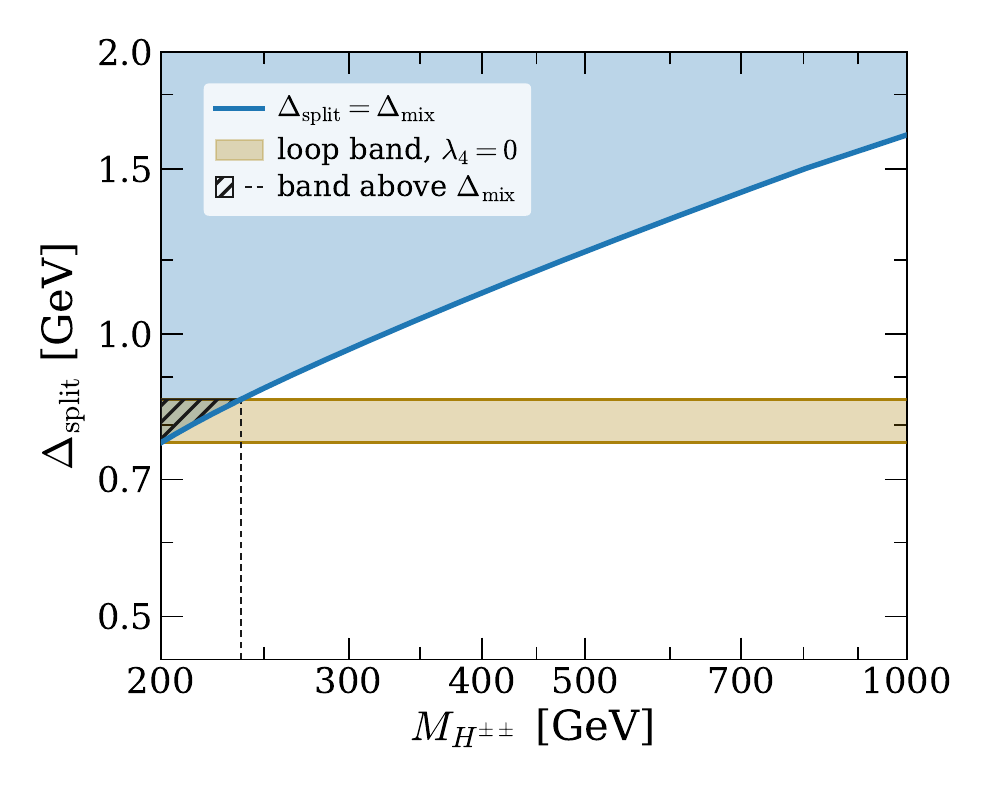}
\caption{Mass splitting against the doubly charged Higgs mass.  Blue solid: $\Delta_{\rm mix}$, the upper edge of the window in the splitting, at the crossover VEV of each mass, drawn as the mean of the two hadronic methods.  Light blue region: splittings at which the cascade has more than halved the mixed topology.  The mixed topology survives in the region below the blue curve.  Ochre band: the loop splitting band of Fig.~\ref{fig:let-r9}, hatched where it lies above $\Delta_{\rm mix}$.  Vertical dashed rule: $M_{H^{\pm\pm}} = 238$~GeV, the right-hand edge of that hatched region.}
\label{fig:let-window}
\end{figure}
%%%%%%%%%%%%%%%%%%%%

At $\lambda_4 = 0$ the entire splitting is of loop origin.  The mixed fraction is then $0.179$ at $M_{H^{\pm\pm}} = 200$~GeV against the ideal $1/2$, or $0.249$ with the one-loop value $767$~MeV in place of $852$~MeV, and $0.47$ at $1$~TeV where the two treatments agree.  The window closes below $201$ -- $238$~GeV, where the loop band crosses $\Delta_{\rm mix}$ in Fig.~\ref{fig:let-window}, the spread being which edge of the band is taken.  The two hadronic methods stay within $0.79\%$ of their mean along $\Delta_{\rm mix}$.  The choice of $\alpha_{\rm em}$ in Eq.~\eqref{eq:let-closed} moves that mass further, from $224$ to $254$~GeV as the Thomson value or $\alpha_{\rm em}(M_Z)$ replaces the adopted one.

Every number here depends on the assumed neutrino spectrum, and the dependence runs one way.  The cascade width has no neutrino dependence.  At the crossover VEV, the primary two decay rates are equal, so that the denominator of $c/(1-c) = \Gamma_{\rm casc}/(\Gamma_{\ell\ell}+\Gamma_{WW})$ is $2\Gamma_{\ell\ell} = 2\sqrt{AC}$, of which only $A = M_{H^{\pm\pm}}\mathcal{T}/16\pi$ carries the trace.  Hence, $c/(1-c) \propto \mathcal{T}^{-1/2}$, and every effect here grows as $\mathcal{T}$ falls.  Our example of $\sum_i m_i = 0.120$~eV sits at the Planck 2018 limit of $0.12$~eV at $95\%$ confidence~\cite{Planck:2018vyg}, and is therefore the most conservative choice.  Later data are about twice as tight, $\sum_i m_i < 0.071$~eV from the Dark Energy Spectroscopic Instrument (DESI) full shape and baryon acoustic oscillation analysis combined with the cosmic microwave background~\cite{DESI:2024hhd}.  On that lighter spectrum, a lightest neutrino mass of $0.008$~eV giving $\sum_i m_i = 0.071$~eV and $\mathcal{T} = 2.80 \times 10^{-3}$~eV$^2$, the cascade takes $48\%$ of the width at $M_{H^{\pm\pm}} = 200$~GeV in place of $40\%$.  The mixed fraction falls to $0.135$, the window closes below $225$ -- $270$~GeV, and the two exclusion boundaries move by $-1.9\%$ and $+34\%$ in place of $-1.4\%$ and $+24.5\%$.

The shortfall of the mixed fraction below $1/2$ at high mass has a second source, the three-body mode $H^{\pm\pm} \to W^\pm \ell^\pm \bar\nu$.  It has the same $v_\Delta^{-2}$ dependence as the leptonic width, thus their ratio is free of $v_\Delta$.  Unlike the cascade fraction, it does not fall away below the crossover VEV.  At $M_{H^{\pm\pm}} = 1$~TeV, it reaches $2.5 \times 10^{-2}$ of the leptonic width and accounts for $0.012$ of the shortfall in the mixed fraction, against the cascade's $0.017$.  The two do not add to the total $0.028$ because both act on the same factor $(1-c)^2$.  The cascade stays the larger up to about $1.1$~TeV.

%%%%%%%%%%%%%%%%%%%%%%%%%%%%%%%%%%%%%%%%%%%%%%%%%%
\section{Corrections to search limits}
\label{sec:limits}
%%%%%%%%%%%%%%%%%%%%%%%%%%%%%%%%%%%%%%%%%%%%%%%%%%

An exclusion boundary is where the predicted rate meets the measured upper limit, $\sigma_{\rm DY} \,{\rm BR}^2 = \sigma_{95}$.  Both cross sections depend on $M_{H^{\pm\pm}}$ alone, so at fixed mass the boundary is the locus ${\rm BR} = \sqrt{\sigma_{95}/\sigma_{\rm DY}}$.  The cascade lowers ${\rm BR}$ at fixed $v_\Delta$, so the boundary moves until that value is restored.  Differentiating with respect to $\log v_\Delta$ at fixed cascade width, with ${\rm d}\Gamma/{\rm d}\log v_\Delta = 2\left(\Gamma_{WW} - \Gamma_{\ell\ell}\right)$, gives
\begin{equation}
\begin{split}
\frac{{\rm d}\log {\rm BR}(WW)}{{\rm d}\log v_\Delta}
&= 2\left[ 1 - {\rm BR}(WW) + {\rm BR}(\ell\ell) \right]
~, \\
\frac{{\rm d}\log {\rm BR}(\ell\ell)}{{\rm d}\log v_\Delta}
&= -2\left[ 1 + {\rm BR}(WW) - {\rm BR}(\ell\ell) \right]
~,
\end{split}
\label{eq:let-slope}
\end{equation}
With ${\rm BR}(WW) \to 1$ and ${\rm BR}(\ell\ell) \to 0$ the slopes become $0$ and $-4$.  The dominant mode no longer responds to $v_\Delta$, while the other responds at full power.  There ${\rm BR}(\ell\ell) \propto v_\Delta^{-4}$, and the observed rate carries ${\rm BR}^2$, so the signal goes as $v_\Delta^{-8}$.  Restoring a lost signal factor $f$ then costs a shift $f^{-1/8}$: the boundary moves as the eighth root of the signal.

The above counting treats the cascade fraction as fixed, so it holds only while the cascade is a small part of the width.  For a large cascade, the exact statement is as follows.  Holding (i) ${\rm BR}(\ell\ell)$ or (ii) ${\rm BR}(WW)$ fixed is the same as holding $\left(\Gamma_{WW} + \Gamma_{\rm casc}\right)/\Gamma_{\ell\ell}$ or $\Gamma_{WW}/(\Gamma_{\ell\ell} + \Gamma_{\rm casc})$ fixed, respectively, both quadratic in $v_\Delta^2$ with a single positive root.  Evaluated at the unshifted boundary of $v_\Delta$, the shifted $v_\Delta'$ is given by
\begin{equation}
\frac{v_\Delta'}{v_\Delta} 
= 
\left(\frac{\sqrt{\kappa^2+4} \mp \kappa}{2}\right)^{\!1/2}
~,
\label{eq:let-shift}
\end{equation}
where one should take $\kappa = \kappa_{\ell\ell} \equiv \Gamma_{\rm casc}/\Gamma_{WW}$ and the upper sign in case (i) or $\kappa_{WW} \equiv \Gamma_{\rm casc}/\Gamma_{\ell\ell}$ and the lower sign in case (ii).

\begin{figure}[tbp]
\centering
\includegraphics[width=1.05\columnwidth]{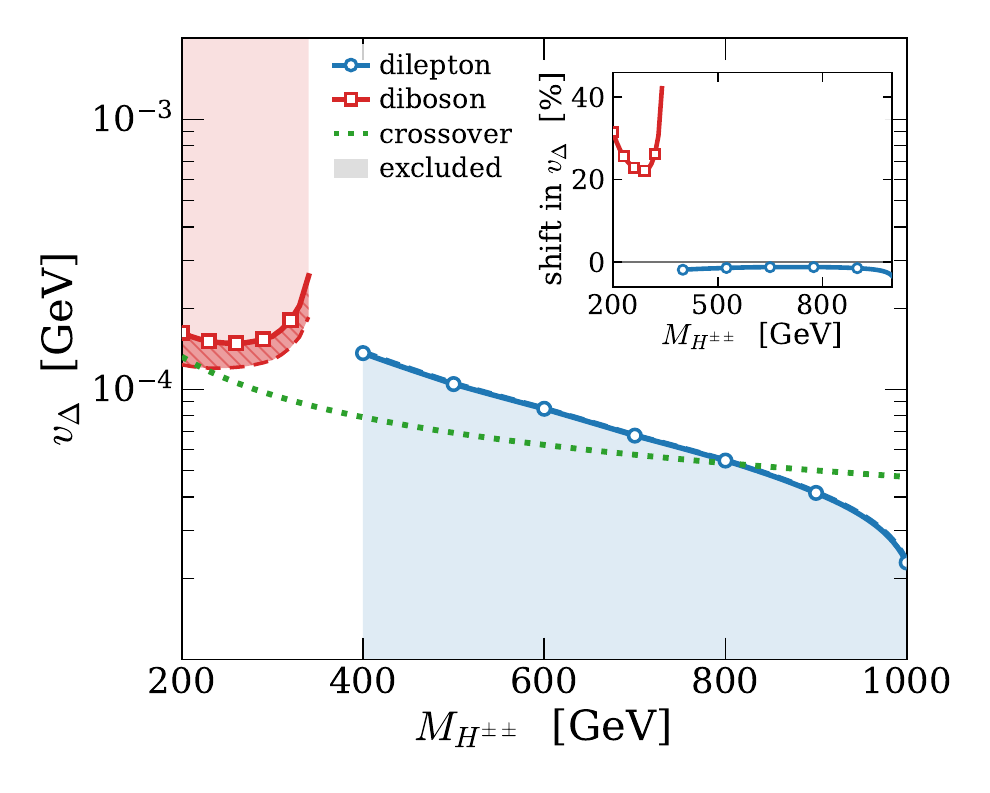}
\caption{Triplet VEV against the doubly charged Higgs mass, from the published ATLAS dilepton (blue circles over $400$ to $1000$~GeV) and diboson (red squares over $200$ to $340$~GeV) limits at $139$~fb$^{-1}$.  The solid curves take into account the cascade at the reference splitting $\Delta_{\rm split} = 852$~MeV, applied at every mass.  The dashed curves are for zero splitting, same as the ATLAS results.  The light shaded regions are excluded with the cascade decay included.  The darker strips mark the differences between our corrected boundaries and the ATLAS ones, narrower than the line itself on the dilepton side.  The dotted green curve marks where the leptonic and diboson widths are equal.  Inset: the fractional change of each boundary against mass.}
\label{fig:let-excl}
\end{figure}

Both boundaries lie above the crossover VEV, where $\Gamma_{WW}$ is larger than $\Gamma_{\ell\ell}$.  At $\Delta_{\rm split} = 852$~MeV, the median $\kappa_{\ell\ell}$ over $400$ to $1000$~GeV is $0.055$, so Eq.~\eqref{eq:let-shift} with the upper sign gives the correction $-1.4\%$.
The diboson limit is currently set between $M_{H^{\pm\pm}} = 200$ and $340$~GeV, where the required ${\rm BR}(WW)$ runs from $0.4306$ to $0.9524$.  That sends the first line of Eq.~\eqref{eq:let-slope} from $1.52$ down to $0.118$, about $2.5$ to $34$ times below 4, so the same loss of signal needs a much larger move in $v_\Delta$.  There $\kappa_{WW}$ lies between $0.83$ and $1.54$, an order of magnitude above the dilepton side.  At $M_{H^{\pm\pm}} = 240$~GeV, where the diboson shift takes its median value and $\kappa_{WW} = 0.904$, the correction is $+24.5\%$.  The same channel therefore moves the two boundaries by $-1.4\%$ and $+24.5\%$.  Both figures are reinterpretations at the level of branching ratios, computed at fixed experimental acceptance.  A full recast would have to allow for the soft hadrons and leptons added by the cascade decay to the event, which can change the acceptance of a diboson selection in particular, where the cascade products fall near the objects the selection is built on.  Across their mass ranges, the dilepton shift stays between $-1.2$ and $-3.4\%$ while the diboson one runs from $+22.1$ to $+42.1\%$, as Fig.~\ref{fig:let-excl} shows.

Both searches lose reach, leaving a wider uncovered range of $v_\Delta$ at each mass than the published translations show~\cite{Ashanujjaman:2021txz,Bolton:2024thn,Antusch:2018svb,BhupalDev:2018tox,Han:2015hba,Kanemura:2014goa}.  No collaboration search targets the mixed final state.  Each ATLAS analysis quotes its limit for a doubly charged Higgs boson decaying entirely into its own channel, ${\rm BR}(\ell^\pm\ell^\pm) = 1$ for the dilepton search and ${\rm BR}(W^\pm W^\pm) = 1$ for the diboson one~\cite{ATLAS:2022pbd,ATLAS:2021jol}.  At the crossover VEV, however, the mixed topology is the largest of the three: $0.434$ against $0.217$ for each pure topology at $M_{H^{\pm\pm}} = 500$~GeV with the cascade included.  An event with two leptons and four jets has almost no acceptance in a four-lepton selection.  Over two fifths of the sample is therefore collected by neither search.  The only limits on it are a recasting of published efficiencies~\cite{delAguila:2013mia} and a projection~\cite{Bolton:2024thn}.

%%%%%%%%%%%%%%%%%%%%
%%%%%%%%%%%%%%%%%%%%

More data narrows the uncovered region and, over one range of masses, closes it.  The significance grows as the square root of the integrated luminosity $L_{\rm int}$, which improves the excludable signal strength as $L_{\rm int}^{-1/2}$.  This is a statistics only projection.  It holds the systematic uncertainties and the selection fixed, and either of those may dominate the gain by $3000$~fb$^{-1}$.  Between $139$ and $3000$~fb$^{-1}$ the dilepton boundary rises by a factor $1.24$ at $M_{H^{\pm\pm}} = 400$~GeV to $1.39$ at $850$~GeV, in comparison with the $(3000/139)^{1/16} = 1.21$ of the eighth-root rule.  Above $850$~GeV, the factor grows quickly, reaching $2.21$ at $1$~TeV.  The boundary has fallen below the crossover VEV by then, so the leptonic branching ratio is near one and its response to $v_\Delta$ has collapsed, which is the same mechanism that governs the diboson side below.

The diboson boundary sits where the response of Eq.~\eqref{eq:let-slope} is small.  Hence, the same gain in sensitivity produces a much larger displacement, downward in $v_\Delta$.  It falls by a factor of $1.5$ at $200$~GeV to $3.0$ at $340$~GeV, with the cascade at the reference splitting.  The displacement is self-limiting.  As the boundary descends, ${\rm BR}(WW)$ falls from $0.95$ to $0.44$ and the response of Eq.~\eqref{eq:let-slope} steepens from $0.12$ to $1.9$, with the boundary coming to rest near the crossover VEV.  At $139$~fb$^{-1}$, no mass has both boundaries.  At $3000$~fb$^{-1}$, the diboson search can reach $550$~GeV at branching ratio one.  From $400$ to about $519$~GeV, its boundary will lie below the dilepton one, leaving no gap in $v_\Delta$.

%%%%%%%%%%%%%%%%%%%%%%%%%%%%%%%%%%%%%%%%%%%%%%%%%%
\section{Summary}
\label{sec:summary}
%%%%%%%%%%%%%%%%%%%%%%%%%%%%%%%%%%%%%%%%%%%%%%%%%%

In this work, we discuss phenomenological consequences of the doubly charged Higgs search at the LHC when the mass splitting among the triplet states is in the sub-GeV regime.  First, whether the mixed same-sign topology exists is a condition on $\lambda_4$ rather than on the splitting.  At $M_{H^{\pm\pm}} = 200$~GeV, it requires $\lambda_4 > 4.1 \times 10^{-5}$, which is $223$ times smaller than the shift in $\lambda_4$ produced by a change of the renormalization scale by a factor of ten.

Thus, at that mass the answer depends on the scale at which $\lambda_4$ is defined.  Since $\lambda_4(\overline\mu = M_{H^{\pm\pm}})$ is a parameter of the potential like any other, the condition is definite once the scale is fixed.  Above $\simeq 200$~GeV, the question does not arise at all because $\lambda_4^{\rm open}$ is negative there and the mixed topology survives from $\lambda_4 = 0$ up to $\lambda_4^{\rm cancel}$.

Second, we compute the cascade decay rate from the measured spectral functions.  The rate built from hadronic data is $0.684 \pm 0.022$ ($4.87 \pm 0.09$) of the partonic rate at $852$~MeV ($200$~MeV).  Two calculations of that rate are carried out, and the quoted band covers both of them together with the systematic uncertainty each one has, which is why it is not narrowed by averaging their difference away.  They are not two independent measurements.  Below about $500$~MeV, both are the single pion pole built from the same decay constant, where they agree to $0.35\%$.  The growth of the cascade rate with the mass splitting does not follow the fifth power law given in Eq.~\eqref{eq:let-fifth}.

Third, a dilepton boundary sits where the cascade is small in comparison with the diboson width and, thus, it barely moves.  At the lower masses of its range, the leptonic branching ratio is still falling as $v_\Delta^{-4}$.  A diboson boundary sits where its own branching ratio is near one and cannot respond, and where the cascade is measured against the much smaller $\Gamma_{\ell\ell}$.  The same channel therefore shifts the two by $-1.4\%$ and $+24.5\%$.

No collaboration analysis targets the mixed final state, in which one $H^{\pm\pm}$ gives two leptons and the other gives four jets through two $W$ bosons.  At the crossover VEV that topology has the largest branching ratio of the three, $0.434$ against $0.217$ for each pure one at $M_{H^{\pm\pm}} = 500$~GeV, and it has almost no acceptance in a four-lepton selection.  Over $2/5$ of the sample is therefore collected by neither search, and the published translations of the two limits into $v_\Delta$ are correspondingly too strong.

At $139$~fb$^{-1}$ the range of $v_\Delta$ left uncovered is open on one side at every mass, since no mass carries both boundaries.  With a naive scaling up to $3000$~fb$^{-1}$, the diboson search can reach $550$~GeV at branching ratio one, and from $400$ to about $519$~GeV its boundary lies below the dilepton one, leaving no gap.  A dedicated selection for two leptons together with four jets would recover the part of the sample that a sub-GeV splitting moves out of both existing analyses.

The splitting is fixed by the potential up to one coupling, not a nuisance parameter to be scanned over, and its natural value places the spectrum where neither search is calibrated.  Degeneracy means a splitting below the few MeV where the cascade becomes negligible compared to the other two channels.  By Eq.~\eqref{eq:let-family}, that is a coupling window of order $10^{-4}$ against the $0.020$ of $\lambda_4^{\rm cancel}$.  Therefore, a search designed at degeneracy is designed at a point that requires $\lambda_4$ tuned to $\lambda_4^{\rm cancel}$ within about $1\%$.

Below about $M_{H^{\pm\pm}} = 300$~GeV, an exclusion boundary drawn at one splitting is a boundary drawn at one value of $\lambda_4$, and that value matters more than any experimental uncertainty.  A limit in the plane of mass against $v_\Delta$ should therefore state at which $\lambda_4$, and at which scale, it was drawn.  Better still is the plane of mass against $\lambda_4$ of Fig.~\ref{fig:let-lambda4}, which shows what part of the coupling range a conclusion covers.

\begin{acknowledgments}
The author thanks K.~Yagyu for his hospitality during his visit at Tokyo University of Science, where this work was initiated.  This work was supported in part by the National Science and Technology Council under Grant No.~NSTC114-2112-M-002-020-MY3.  Numerical calculations, including coding and plots, and part of the drafting, were done with the assistance of Claude Opus 5.  The author takes full responsibility for the work.
\end{acknowledgments}

%%%%%%%%%%%%%%%%%%%%%%%%%%%%%%%%%%%%%%%%%%%%%%%%%%
\appendix
%%%%%%%%%%%%%%%%%%%%%%%%%%%%%%%%%%%%%%%%%%%%%%%%%%

%%%%%%%%%%%%%%%%%%%%%%%%%%%%%%%%%%%%%%%%%%%%%%%%%%
\section{One-loop splitting of a scalar multiplet}
\label{sec:app-loop}
%%%%%%%%%%%%%%%%%%%%%%%%%%%%%%%%%%%%%%%%%%%%%%%%%%

Suppose the multiplet is a complex scalar of weak isospin $T$ and hypercharge $Y$, the triplet of interest having $T = 1$ and $Y = 2$.  The one-loop gauge splitting of such a multiplet is already given in Ref.~\cite{Cirelli:2005uq} for arbitrary $Q$, $Q'$ and $Y$, together with its heavy mass limit and the observation that the residual divergence is absorbed by the quartic coupling of the multiplet to the doublet.  Nothing in it uses the two values above, so it applies as it stands to the larger multiplets proposed for the neutrino mass, which contain states of higher charges~\cite{Ghosh:2018drw,Giarnetti:2023dcr}.  Only the two results used in this work are collected here, in the conventions of Sec.~\ref{sec:model}.

Expanding the kinetic term of Eq.~\eqref{eq:let-kin} to second order in the gauge fields gives the couplings $c_\gamma = eQ$ and $c_Z = (e/s_W c_W)(Q c_W^2 - Y/2)$, and with them the only two diagrams the kinetic term supplies at one loop, the exchange of a gauge boson across the scalar line and the seagull diagram.  Evaluating them on the mass shell, $p^2 = M^2$ with $M$ the common tree-level mass, the pole shift $\delta M_V = \Sigma_V(M^2)/2M$ gives, for the difference between two components of charges $Q$ and $Q'$,
\begin{align}
\begin{split}
M_Q - M_{Q'} = & - \frac{1}{8\pi} \sum_V \delta c_V^2\, m_V \\
& {} + \frac{2 - 3\bar{L}}{32 \pi^2 M_{H^{\pm\pm}}} \sum_V \delta c_V^2\, m_V^2 + \dots
~,
\end{split}
\label{eq:app-leading}
\end{align}
with $\bar{L} \equiv 1/\bar\epsilon + \log(\overline\mu^2/M^2)$ collecting the divergence $1/\bar\epsilon$ and the renormalization scale.  The first term is the closed form of Eq.~\eqref{eq:let-closed}.  The photon does not enter it, since $m_\gamma = 0$.  So the sum runs over the massive mediators only.  What the photon supplies is the condition $\sum_V \delta c_V^2 = 0$, which is what makes the difference of the two self energies finite and which fixes $\delta c_W^2$ in terms of the other two.  The second is smaller by $m_V/(2\pi M_{H^{\pm\pm}})$ and carries the entire renormalization scale dependence.  On the mass shell, the result is independent of the gauge parameter for each component separately.

The difference is not finite, however.  For the $H^{\pm\pm} - H^\pm$ gap of the $T = 1$, $Y = 2$ triplet the sum reduces to a closed form.  With $\delta c_Z^2 = g^2\left(c_W^2 - 2 s_W^2\right)$ and the two $W$'s contributing $-g^2$ between them, and using $M_W^2 = M_Z^2 c_W^2$,
\begin{align}
\sum_V \delta c_V^2\, m_V^2 &= -2 g^2 s_W^2 M_Z^2 
\simeq -1.58 \times 10^{3}~{\rm GeV}^2
~.
\label{eq:app-residue}
\end{align}
The $1/\bar\epsilon$ inside $\bar{L}$ therefore survives in the second term.  It is absorbed by the counterterm of $\lambda_4$.  More explicitly, the residue carries exactly two powers of $v$ and one power each of $g^2$ and $g'^2$, which matches the form of $\lambda_4 \Phi^\dagger\Delta\Delta^\dagger\Phi$.  Furthermore, $\lambda_4$ is the only tree-level coupling that splits the components.  What is finite is the sum of Eq.~\eqref{eq:app-leading} with $-\lambda_4(\overline\mu) v^2/(8M_{H^{\pm\pm}})$, which is Eq.~\eqref{eq:let-family}.

The scale dependence is therefore a shift of $\lambda_4$ alone, Eq.~\eqref{eq:let-loopswing}, and is proportional to $1/M_{H^{\pm\pm}}$ as Table~\ref{tab:loop-scale} shows.  Summing the mediators with their weights, the complete function falls short of Eq.~\eqref{eq:let-closed} by $10.0\%$ at $M_{H^{\pm\pm}} = 200$~GeV and by $1.4\%$ at $1$~TeV.  The unrounded values behind those two percentages are $767.4$ and $840.7$~MeV against the limit $852.3$~MeV.

%%%%%%%%%%%%%%%%%%%%%%%%%%%%%%%%%%%%%%%%%%%%%%%%%%
\section{The two hadronic methods}
\label{sec:app-routes}
%%%%%%%%%%%%%%%%%%%%%%%%%%%%%%%%%%%%%%%%%%%%%%%%%%

The virtual $W$ boson of Eq.~\eqref{eq:let-cascade} has an invariant mass at most $\Delta_{\rm split}$, which is considered to be a few hundred MeV in this work.  Its decay products are therefore hadrons, which the partonic counting of Eq.~\eqref{eq:let-fifth} fails to describe.  What replaces it is the same quantity that describes $\tau$ decay at the same invariant mass, since both proceed through the charged weak current at $q^2 \ll M_W^2$.  The hadronic side of that current is carried by two spectral functions, one for the spin-1 part and the other for the spin-0 part.  The scalar side supplies the derivative coupling of Eq.~\eqref{eq:let-cascade}.  Contracting the two and integrating over the pair's phase space yields the rate as a single integral over $s = q^2$ from zero to $\Delta_{\rm split}^2$.

Writing $M$ for $M_{H^{\pm\pm}}$ and $M' = M - \Delta_{\rm split}$ for the singly charged mass, the rate follows from the same construction that gives the $\tau$ hadronic width in terms of these spectral functions~\cite{Braaten:1991qm,Davier:2013sfa}, with the overall factor fixed as described at the end of this appendix,
\begin{equation}
\Gamma_{\rm had}
= \frac{g^4}{256\pi^3 M^3}\int_0^{\Delta_{\rm split}^2}\!\!{\rm d}s\,
\left[\rho^{(1)} F_1 + \rho^{(0)} F_0\right](s)
~,
\label{eq:app-master}
\end{equation}
with the two weights and the K\"all\'en function $\lambda(a,b,c) = a^2 + b^2 + c^2 - 2(ab + bc + ca)$~\cite{Byckling:1971vca}, here evaluated at $(M^2, M'^2, s)$ and factorized,
\begin{align}
\begin{split}
F_1 &= \frac{\lambda^{3/2}}{D_W}
~, \\
F_0 &= \frac{\lambda^{1/2}}{D_W}\,
\Delta_{\rm split}^2\left(M+M'\right)^2\left(1-\frac{s}{M_W^2}\right)^{\!2}
~, \\
\lambda &= \left(\Delta_{\rm split}^2 - s\right)\left[\left(M+M'\right)^2 - s\right]
~, \\
D_W &= \left(s-M_W^2\right)^2 + M_W^2\Gamma_W^2
~.
\end{split}
\label{eq:app-weights}
\end{align}
The recoil of the $H^\pm$ is kept exactly, which reduces the closed forms of Ref.~\cite{FileviezPerez:2008jbu} by a factor of $1 - 3\Delta_{\rm split}/2M$, which is about $0.64\%$ for $M_{H^{\pm\pm}} = 200$~GeV and the reference splitting.  The spin-1 weight includes $\lambda^{3/2}$, one power of $\lambda^{1/2}$ from the two-body phase space and one power of $\lambda$ from the transverse invariant.  The spin-0 weight has $\lambda^{1/2}$ and survives only because $q\cdot(p+p') = M^2 - M'^2$ does not vanish.

The spectral functions are normalized so that $\rho^{(1)} \to 1$ for three massless colored quark pairs and $1/3$ for one massless lepton pair.  In this convention, the single pion enters $\rho^{(0)}$ as $2\pi^2 |V_{ud}|^2 f_\pi^2\,\delta(s - m_\pi^2)$ and the kaon as $2\pi^2 |V_{us}|^2 f_K^2\,\delta(s - m_K^2)$.  The overall factor is not fitted.  It is fixed by requiring that Eq.~\eqref{eq:app-master}, evaluated with one massless lepton pair in place of the hadrons, return the closed form $g^4\Delta_{\rm split}^5/(240\pi^3 M_W^4)$.

The spin-0 part is not a small correction here.  The single pion channel runs through it, and below the value quoted at the end of this appendix it exceeds the entire partonic rate.  The rate that enters $R$ of Eq.~\eqref{eq:let-Rdef} is $\Gamma_{\rm hadron} = \Gamma_{\rm had} + \sum_\ell \Gamma_\ell$.  The three lepton pairs follow from the same master formula, with $\rho^{(1)}$ and $\rho^{(0)}$ replaced by one free pair and their charged lepton masses kept.  They carry $31\%$ of that numerator at $\Delta_{\rm split} = 852$~MeV, and they also appear in the denominator of $R$.  This is the counting used in Sec.~\ref{sec:hadronic} to fix $\Delta_\times$.

%%%%%%%%%%%%%%%%%%%%%%%%%%%%%%%%%%%%%%%%%%%%%%%%%%
\subsection{First method}
\label{sec:app-first}
%%%%%%%%%%%%%%%%%%%%%%%%%%%%%%%%%%%%%%%%%%%%%%%%%%

Take the spin-1 spectral function from the non-strange vector plus axial vector data extracted from $\tau$ decay~\cite{Davier:2013sfa}, which are measured and require no model of the resonances.  Those data exclude $\tau \to \pi\nu$ from their normalization and therefore do not include a single pion channel.  The spin-0 part is supplied separately as poles, $2\pi^2 |V_{ud}|^2 f_\pi^2\,\delta(s - m_\pi^2)$ and $2\pi^2 |V_{us}|^2 f_K^2\,\delta(s - m_K^2)$.  The pion pole lies below the two-pion threshold and accounts for the entire hadronic rate at splittings of $200$ to $300$~MeV.  From the two-pion threshold to about $\sqrt s = 1$~GeV, the $\rho$ meson dominates.

Above $\sqrt s = 1.55$~GeV, we continue the spin-1 part with the massless perturbative series through order $\alpha_s^3$~\cite{Braaten:1991qm}.  That point lies below the kinematic limit $\sqrt s = m_\tau = 1.777$~GeV of the $\tau$ data.  It is chosen because the measured spectral function and the perturbative series already agree there to a few percent, while the bins above it have rapidly growing uncertainties.

Since the integral in Eq.~\eqref{eq:app-master} runs only to $s = \Delta_{\rm split}^2$, the continuation enters no splitting below $1.55$~GeV and the charm channel closed below the $c\bar d$ threshold at $1.89$~GeV.  Every value of $R$ this paper quotes below $\Delta_{\rm split} = 1.5$~GeV therefore rests on the measured data and the two poles alone.  That the method reproduces the partonic rate at large splitting follows from the normalization fixed above and is not an independent check.

%%%%%%%%%%%%%%%%%%%%%%%%%%%%%%%%%%%%%%%%%%%%%%%%%%
\subsection{Second method}
\label{sec:app-second}
%%%%%%%%%%%%%%%%%%%%%%%%%%%%%%%%%%%%%%%%%%%%%%%%%%

The second method sums exclusive channels.  The single pion, the single $K$, the $D$, and the $D_s$ enter through their decay constants.  The resonances enter through relativistic Breit-Wigner shapes with energy-dependent widths.  The $a_1$ has the three-pion final state, the $K^*$ has $K\pi$, and the four-pion channel is included as well, each normalized to the corresponding measured $\tau$ branching ratio.  The $\rho$ meson has two pions and is normalized instead by its measured partial width into an electron pair, $\Gamma(\rho\to e^+e^-)$.  The three leptonic channels are kept with their charged lepton masses.

What is left out is $2.6\%$ of the total width of the $\tau$ and $4.0\%$ of its hadronic width, and it is included as a single effective channel in the systematic band.  This residue, rather than any breakdown of the method, limits the method to splittings below about $2$~GeV.  Above $\Delta_{\rm split} = 1.87$~GeV, the charged $D$ mass, the adopted mean is therefore continued by holding the ratio of the two methods at its value there rather than by averaging them.  Every splitting quoted in this paper lies below that point and, thus, no result depends on the continuation.

The two methods treat the resonance region differently and share no spectral function, but they are not independent everywhere.  At the splitting of interest here, both methods build the single pion and single $K$ poles from the same two decay constants in the same closed form.  The three leptonic channels are a second closed form they hold in common.  On that basis, the part of the rate on which the two actually differ is $11\%$ of this method's total rate and $14\%$ of the first method's.  Both fractions grow with the splitting, reaching $43$ and $45\%$ at $1.5$~GeV.  The agreement quoted next is therefore a test of the resonance treatment, which is where they differ, and not of the total rate.

The two methods lie within $2.4\%$ of their mean over $200$~MeV $\le \Delta_{\rm split} \le 2$~GeV, which is where both apply, the largest departure falling near the upper end.  The adopted values quoted in Eq.~\eqref{eq:let-R9} are the mean of the two methods.  The uncertainty covers the full envelope of both, that is the spread between the methods together with each method's own systematic band.  That envelope is the wider of the two measures at every splitting quoted, and it is the union of the two shaded bands of Fig.~\ref{fig:let-r9}.  The single pion channel alone exceeds the total partonic rate below $504$~MeV, where $R_\pi = 1$ in Eq.~\eqref{eq:let-Rpi} with the pion mass kept, which is why $R$ rises rather than falls there.

\end{document}